\documentclass[conference]{IEEEtran}
\IEEEoverridecommandlockouts
\usepackage{cite}
\usepackage{amsmath,amssymb,amsfonts}
\usepackage{algorithmic}
\usepackage{graphicx}
\usepackage{textcomp}
\usepackage{xcolor}
\usepackage[hyphens]{url}
\usepackage{soul}
\usepackage{multirow}

\def\BibTeX{{\rm B\kern-.05em{\sc i\kern-.025em b}\kern-.08em
    T\kern-.1667em\lower.7ex\hbox{E}\kern-.125emX}}
    
\begin{document}    
\title{Blockchain-Enabled NextGen Service Architecture for Mobile Internet Offload}

\author{\IEEEauthorblockN{Raman Singh}
\IEEEauthorblockA{\textit{School of Comp Sci \& Stats} \\
\textit{Trinity College Dublin} \\
Dublin, Ireland \\
\textit{Thapar Institute of Engineering \& Technology}\\
Patiala, India \\
raman.singh@thapar.edu}
\and
\IEEEauthorblockN{Hitesh Tewari}
\IEEEauthorblockA{\textit{School of Comp Sci \& Stats} \\
\textit{Trinity College Dublin}\\
Dublin, Ireland \\
htewari@tcd.ie}
}

\maketitle

\section{Abstract}
The amalgamation of different generations of mobile cellular networks around the globe has resulted in diverse data speed experiences for end users. At present there are no defined mechanisms in place for a subscriber of one mobile network operator (MNO) to use the services of a WiFi provider. Cellular and Data Service providers also have no standardized procedures to securely interact with each other, and to allow their subscribers to use third party services on a pay-as-you-go basis. This paper proposes a blockchain-based offloading framework that allows a subscriber of a mobile operator to temporarily use another MNO or WiFi provider’s higher speed network. Smart contracts allow diverse entities such as MNOs, Brokers and WiFi Providers to automatically execute mutual agreements to enable the utilization of third party infrastructure in a secure and controlled manner. To test the proposed framework, the offloading of a subscriber from 3G/4G/4G-LTE/5G networks to a fixed broadband WiFi network was carried out and the results analyzed. The offloading framework was implemented using the ns-3 network simulator, and the Ethereum blockchain smart contract features were used for the settlement of invoices.

\section{Introduction}
The global rollout of 5G networks is now gathering momentum, as more and more countries are deploying this state-of-the-art broadband cellular network technology. However, at the same time many countries still have operational legacy mobile networks such as 4G, 4G-LTE, or even 3G. Even in the places where 5G networks are available, coverage is not always universal, and many pockets exist that still run older generation networks (e.g. 2G/3G/4G). A report published by the GSM Association (GSMA) \cite{gsmareport2020} suggests that at the end of 2019, 4G coverage was about 50\% of the total mobile Internet availability by geographical area. Due to several advantages such as lower cost and fixed broadband/fiber infrastructure, WiFi still provides higher bandwidth speeds to its users, and is popular among small and big organizations and also in retail and residential settings.

Today the world is using mobile cellular technologies such as 3G, 4G, 4G-LTE and 5G with varying data transfer speeds. We believe that an improved user experience can be gained by offloading such cellular network users to local, higher speed networks. For example, if WiFi providers could allow mobile subscribers to use their fixed broadband infrastructure and in return get monetary reward for their services from a MNO, then one can reduce the load on the cellular networks and simultaneously increase the data speed offered to users. The proposed framework allows independent private WiFi operators to be paid for their services by offloading users onto their networks, and that in turn means less capital expenditure investment by the MNOs. Some studies like the one released by OpenSignal \cite{opensignal} suggest that in the future, mobile Internet will transcend WiFi speeds, then offloading can also be performed from WiFi to mobile network infrastructure. The second reason for offloading can be the guarantee of services to subscribers by the MNO where they do not have a license to operate, or have poor signal coverage issues. These subscribers can be offloaded to partner WiFi providers and the subscriber will be able to enjoy enhanced data speeds. The third rationale for offloading is to ensure better service while roaming. For example, a subscriber who does not have roaming enabled on their device, but wants to use the Internet for short periods of time can be offloaded to one of these high-speed networks.

\begin{figure*}[ht]
\centering
\includegraphics[width=5.3in]{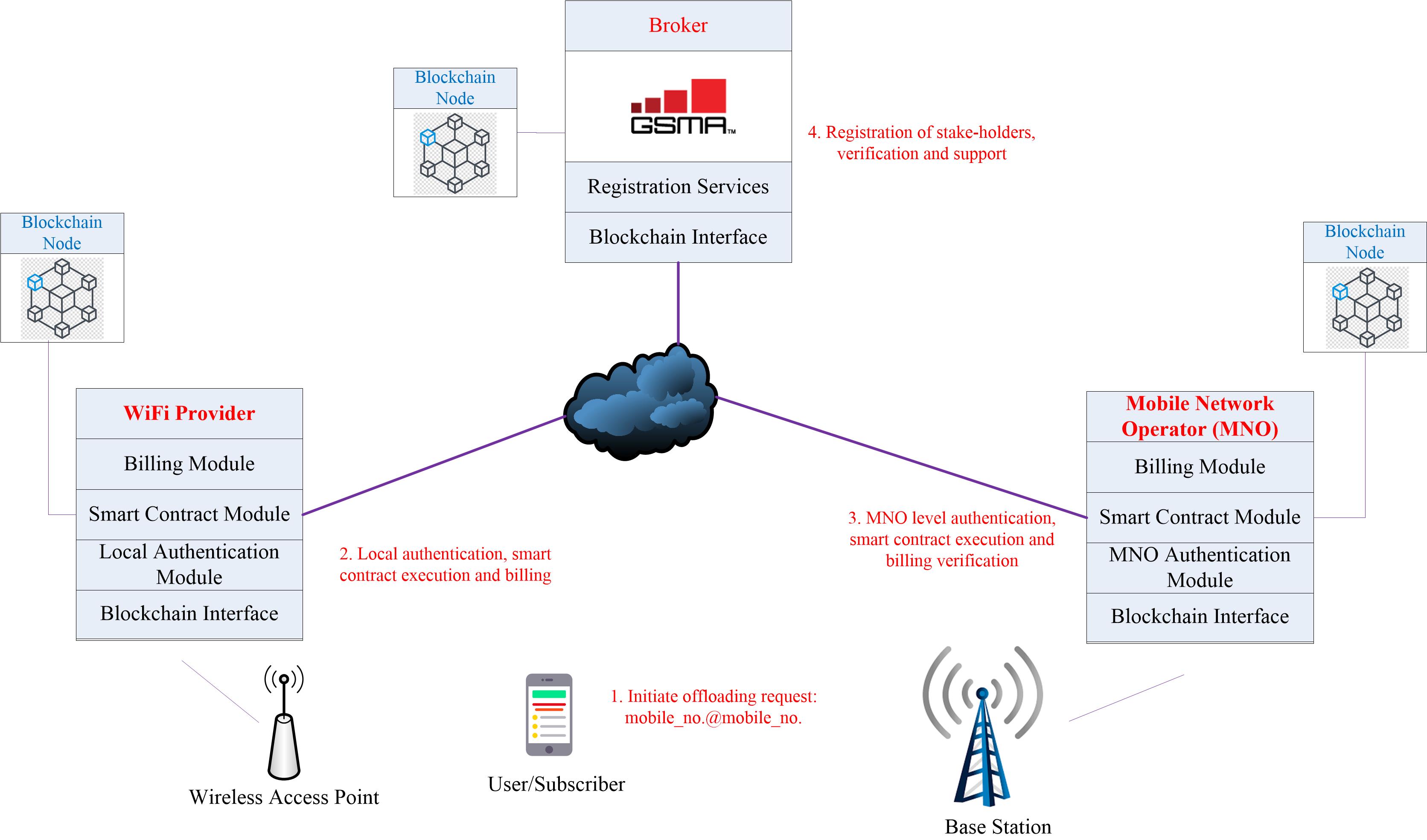}
\caption{System architecture of the proposed framework}
\label{fig:figure1}
\end{figure*}

To allow subscriber roaming between operators and the settlement of usage charges, MNOs at present have memoranda of understanding (MoUs) drawn up between them on a bilateral basis. However, MoUs are complex agreements which take time to negotiate, and therefore it would not be practical to negotiate MoUs with large numbers of small and medium sized WiFi providers on a per MNO basis. To enable the offloading process, MNOs can register with an intermediary/broker who can collectively negotiate MoUs with many WiFi providers on their behalf. Additionally, blockchain technologies can be used to allow the participating  entities to trust each other by executing smart contracts on a blockchain. This append-only distributed ledger technology (DLT), along with a consensus mechanism allows the implementation of smart contracts in real terms, and also digitally facilitates, verifies and enforces the contract between two or more parties \cite{smartcontractpaper}. Smart contracts can act as a bridging gap between stake-holders, and can provide subscribers with a new level of user experience.  

\section{Blockchain-based Subscriber Offloading Framework}
Our proposed offloading framework enables subscribers to temporarily offload their data usage from low-bandwidth to high-bandwidth channels, without changing their network operator. Fig. \ref{fig:figure1} represents the block diagram of the framework and consists of three primary entities, namely a Broker, MNOs and WiFi Providers. The Broker as the name suggests is the ingress of the whole process and is able to coordinate activities between all the entities, as every entity in the system is registered with the Broker. We believe that a GSMA \cite{gsmaofficial} like entity aptly fits the role of the Broker in our proposed system, and all MNOs must register with it. The Broker maintains a blockchain node along with a registration service. The MNO registration process includes setting up a blockchain node for storing smart contracts and transactional data. Organizations that wish to allow their high-speed wireless infrastructure to be used by MNO subscribers must also register with the Broker, along with setting up their corresponding blockchain node. The Broker has oversight during the settlement phase, and also acts as a mediator in the case of any disputes.

The second set of entities in the system are the MNOs that allow their subscribers to opt for offloading to a higher speed network. Reasons for offloading can include low-quality signal coverage, high-speed requirements, roaming to non-serviced areas, or even accessing services provided by particular Data Service providers. This entity includes various functionalities like a blockchain interface, authentication module, smart contract module, and a billing module. The blockchain interface allows various other blockchain nodes to interact with each other, and to update the ledger periodically, including, adding or executing new smart contracts or transactions. The MNO authentication module helps to identify and authenticate a subscriber from the MNOs subscriber database. Since there are many MNOs which are registered with the Broker, and the subscriber should be an active user of that particular MNO, this module identifies a subscriber from an open smart contract, and authenticates its status in order to verify that the user a valid subscriber and is authorized to offload.
 
The smart contract module interacts with open smart contracts to identify users, and sets the values of the parameters in the contracts based on the authentication status, such as success or failure. This module can access the smart contract's data based on the authorization allowed and helps in executing it. The billing module can access the transactions stored in the blockchain, and create a bill based on the executed smart contract which involves a particular MNO. This module then matches the billing amount from the invoice received from the WiFi Provider and authorizes the payment.

\begin{figure*}[t]
\centering
\includegraphics[width=\textwidth]{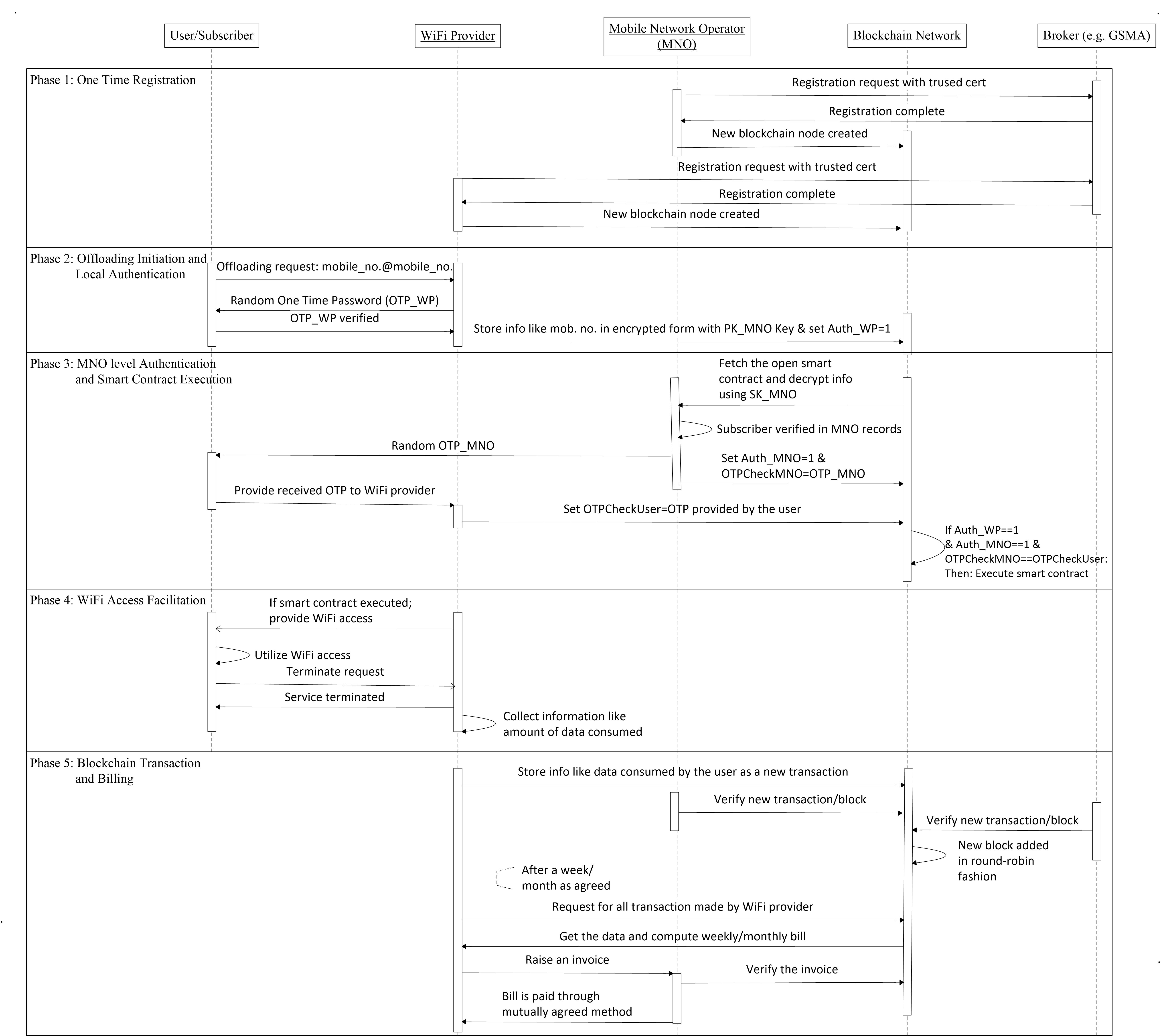}
\caption{Sequence Diagram Illustrating the Various Phases}
\label{fig:figure2}
\end{figure*}

The third set of entities are the WiFi Providers which open up their infrastructure in a controlled manner to the subscribers of MNOs. To expedite the offloading process, a WiFi Provider maintains a blockchain node on its premises. This entity also operates on four modules, such as the blockchain interface, authentication module, smart contract module and billing module. The blockchain interface is responsible for maintaining an up-to-date smart contract and transaction data, along with the full blockchain. The local authentication module ensures the mobile number ownership of the subscriber by validating a one-time password (OTP). This local authentication of the mobile number also avoids the spamming of mobile users or blockchain data. For example, if the local authentication of a mobile number is not concluded successfully, spammers may create millions of offload requests using random mobile numbers, which in turn would create a corresponding number of open smart contracts, and force denial of service attacks against legitimate users. The smart contract module allows the WiFi Provider to create a new smart contract and set its variables based on the subscriber authentication mechanism. Once the subscriber is allowed to offload and subsequently terminates its connection, the billing module records the data usage and writes this as a new transaction to the blockchain. Fig. \ref{fig:figure2} details the various phases of the offloading framework.

\textbf{Phase 1 - One-time Registration:} 
The registration process for a new entity such as a MNO or WiFi Provider wishing to join the system commences with the submission of their trusted third party (TTP) issued public key certificate to the Broker. The Broker which maintains its own blockchain node stores the certificate as a transaction on the blockchain. All the communication amongst the entities in the system is carried out using public-key cryptography, and backed by transparent logs to ensure auditing at a later stage \cite{certauth}. MNOs and WiFi Providers join the blockchain network by creating their corresponding blockchain nodes.

\textbf{Phase 2 - Offloading Initiation and Local Authentication:}
In phase 2, an offloading request is initiated by a user/subscriber. Once the subscriber is in the range of a WiFi Provider and wants to request an offload onto their network, the subscriber connects with the wireless access point (WAP) and accesses the landing page of the WiFi Provider. The landing page can be common for internal and external users. Providing their mobile number in both the username and password fields indicates that the user is external and wishes to initiate an offload procedure. The user also selects its associated MNO from the drop-down menu provided on the landing page. The WiFi Provider generates the random one-time password (OTP\_WP) and sends it to the mobile number entered into the landing page by the user. The user then enters OTP\_WP on the next field of the landing page. Once the WiFi Provider tests the validity of the mobile number, it generates a new smart contract and sets the value of Auth\_WP as 1. The WiFi Provider then encrypts the mobile number of the user with the public key (PK\_MNO) of the mobile operator which it obtains from the blockchain, and assigns this encrypted value to the field given in the smart contract. The WiFi Provider also includes the identity of the concerned MNO in the smart contract, so that all other entities know for whom the smart contract is intended. 

\textbf{Phase 3 - MNO Level Authentication and Smart Contract Execution:}
Phase 3 describes the process undertaken by a MNO. The open smart contracts stored on blockchain are searched by the MNO. Once it finds the intended contract by matching the subscriber’s operator name, it will fetch the encrypted identity of the subscriber and decrypt it using its private key. The decrypted data reveals the mobile number of the subscriber. The MNO tries to verify the identity of the subscriber against its subscriber database, and if successful sets the value of Auth\_MNO to 1 in the smart contract. The MNO also generates a random one-time password (OTP\_MNO) and sets the value of OTPCheckMNO to OTP\_MNO in the smart contract. OTP\_MNO is also forwarded to the subscriber’s mobile number for further processing. Once the subscriber receives the OTP\_MNO value, it enters it on the landing page of the WiFi Provider. The WiFi Provider in turn assigns the OTP\_MNO value to the OTPCheckUser field of the smart contract. Now, if the values of Auth\_WP and Auth\_MNO are both equal to 1, it deduces that both the WiFi Provider and MNO have validated the subscriber’s identity. If the OTPCheckMNO and OTPCheckUser are the same, it represents that the subscriber is authorized to offload, and the smart contract has been executed on the blockchain. The smart contract can also include other information like total time allowed to offload, or any other conditions with the associated offload which need to be honored by all the parties.

\textbf{Phase 4 - WiFi Access Facilitation:}
In phase 4, the WiFi Provider checks the status of the smart contract. If the contract is executed, the WiFi Provider allows access of its services to the subscriber provided by its infrastructure. Primarily this service is high-speed Internet, but it can also be a wide range of other services offered by WiFi providers. When the subscriber  terminates the connection, the WiFi Provider records the data consumed by the subscriber in its logs.

\textbf{Phase 5 - Blockchain Transactions and Billing:}
Phase 5 deals with the transactions and billing-related procedures. The WiFi Provider creates a new transaction in relation to the data consumed by the subscriber and broadcasts it to the blockchain network. This transaction will be verified by all other blockchain nodes, and once verified, will be added to a block by utilizing the proof-of-authority (PoA) consensus mechanism by one of the authorized entities in the system. The PoA algorithm is able to provide faster transaction throughput using a ``identity-as-a-stake" based consensus mechanism. It significantly increases the speed of validating the transactions by generating blocks in a predictable sequence, and hence achieves a better transaction rate when compared with PoW or PoS. The invoice can be generated by the WiFi Provider after an agreed period of time such as a week or a month. The WiFi Provider will access all the transactions made by it from the blockchain data and prepare an invoice based on the mutually agreed per unit price. This invoice will also be verified by the associated MNO and can also be ratified by the broker. Once all parties verify the invoice, the bill will be paid using a mutually agreed out-of-bounds payment mechanism.       

The Ethereum blockchain supports the scalability for large scale offloading requests/transactions using a combination of sharding and side-chains. To test the performance of the Ethereum blockchain, researchers measured 4 million transactions for 380 hours \cite{block_performance}. The experiment concluded that throughput decreases whereas latency increases linearly if we increase the block period which is fixed as per the difficulty level of proof-of-work (PoW). As in our proposed framework, we are using PoA strategy, this bottleneck should not affect the overall throughput and latency of the proposed system. To decrease the time required to complete the workload, the study suggests that powerful machines with high memory and CPUs should be used as blockchain node in PoA mode, for example, the computation time for workload can be reduced by 25\% if the memory is increased from 4GB to 24GB. If the network size is considered, it is found that in 90\%-100\% of cases, matches are found for smaller network sizes, whereas the match ratio is merely 60\%-75\% for larger networks.  
  
\section{Case Studies and Result Analysis}
The proposed framework was implemented using the ns-3 network simulator \cite{ns3official}. Various ns-3 nodes were created to simulate the different entities e.g. Subscriber, MNO, WiFi Provider, Broker and a Data Server. To implement the blockchain functionality, the Ethereum \cite{ethereumofficial} blockchain is implemented on the Docker \cite{dockerofficial} platform. Each node in the ns-3 network is connected to a Docker container using the tap-bridge arrangement of ns-3 \cite{dockerconnection}. In the simulation environment, the subscriber is initially connected to the MNO node, and when the smart contract is executed the connection is switched over to the WiFi Provider. The simulation was carried out for 350 seconds on an Ubuntu Linux based computer running a virtual machine with 8GB RAM, Intel i5 2.50 GHz processor, and 100 GB of allocated memory.

The experimentation was conducted over five separate case studies. The first case study takes the global average of Internet speed and latency for fixed broadband and mobile Internet. The bandwidth given for the fixed broadband is assigned to the WiFi Provider link, whereas the bandwidth given for the mobile Internet is assigned to the MNO link.  In this case study, the subscriber is offloaded from mobile Internet to fixed broadband as per the speed suggested by the global average. In the subsequent case studies, the subscriber is offloaded from 3G, 4G, 4G-LTE, and 5G mobile Internet to fixed broadband i.e. the WiFi Provider. The various data transfer speeds and latencies considered for all case studies are provided in Table \ref{table:results}. 

\begin{table}[ht]
\centering
\caption{Case Studies and Associated Parameters}
\resizebox{\columnwidth}{!}{%
\begin{tabular}{ccccc} \\ \hline
Case Studies & Network Generation & Average Speed (Mbps) & Latency (ms) & Reference \\  \hline
\multirow{2}{*}{\parbox{2cm}{Case Study 1: Global Average}} & Fixed Broadband & 92 & 21 & \cite{speedtest} \\ 
   & Mobile Internet & 46 & 36 & \cite{speedtest}  \\ \hline
\multirow{5}{*}{\parbox{2cm}{Case Study 2: Comparative Average }} & Fixed Broadband & 241 &	13 &	\cite{speedtest} \\ 
& 5G &	71 &	20 &	\cite{speedtest} \\
& 4G-LTE &	50 &	50 &	\cite{edn} \\ 
& 4G &	10 &	100	& \cite{edn} \\
& 3G &	1.5 &	500 &	\cite{edn} 
\\  \hline   
\end{tabular}
}
\label{table:results}
\end{table}

\begin{figure}[ht]
\centering
\includegraphics[width=3in]{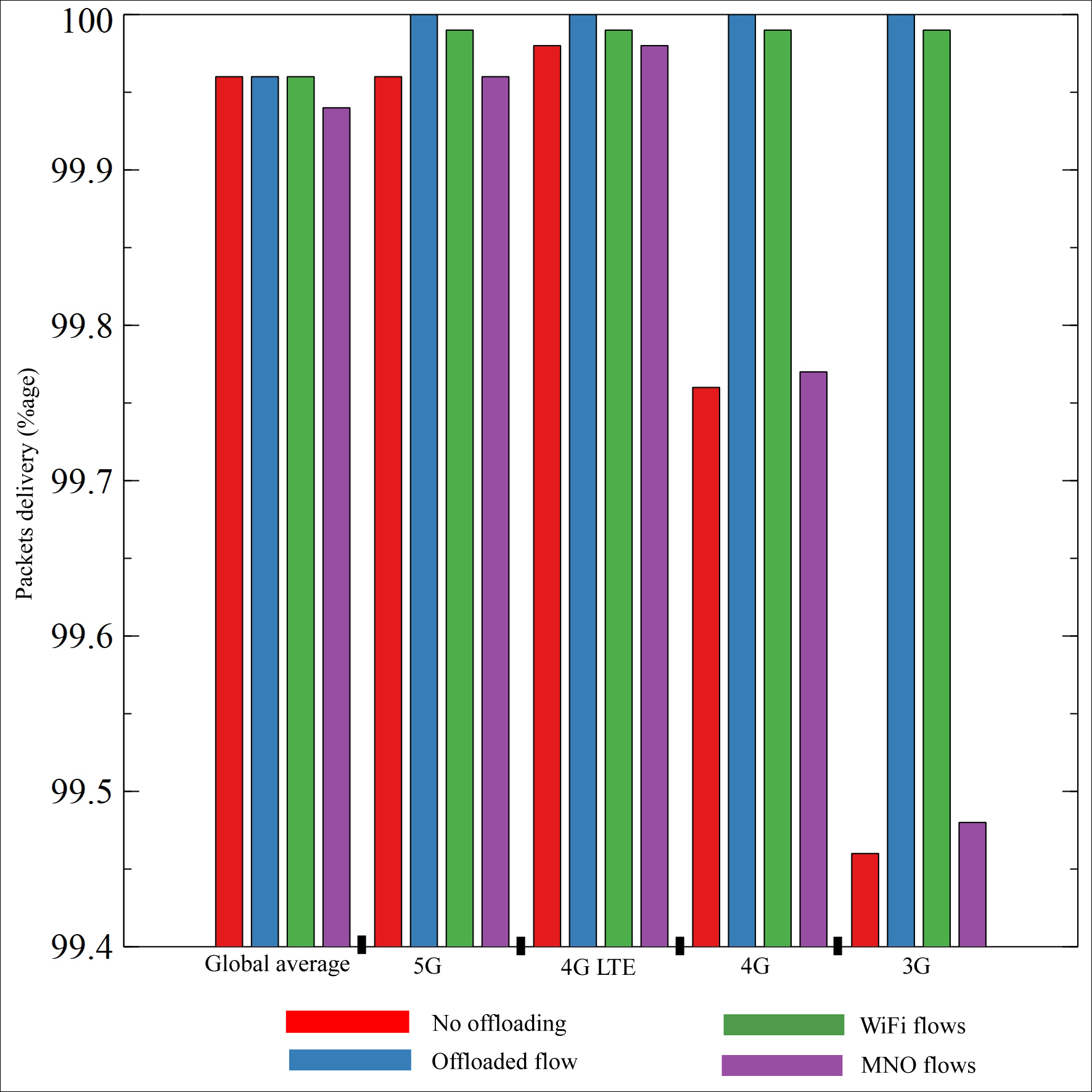}
\caption{Packet Delivery Analysis for Various Types of Flows}
\label{fig:figure3}
\end{figure}

Fig. \ref{fig:figure3} shows the packet delivery percentages for all case studies. The packet delivery is analyzed for all types of flows, such as when the subscriber's packets are not offloaded, for offloaded packets, for packets transmitted through the WiFi link, and other packets of different users which are being transmitted through MNO network. 

For the global average case, the packet delivery percentage is the same for all cases except for the MNO flows, however this difference is insignificant. In this case, 99.96\% of packets are delivered for non-offloaded flows, offloaded flows, and WiFi flows, whereas 99.94\% of total packets are delivered for the MNO flows. For all other case studies, it is evident from Fig. \ref{fig:figure3} that 100\% of offloaded flows are delivered, primarily because of the enhanced data speed of the WiFi link. A slightly less number of packets are delivered in the global average case study if compared to all other case studies, because the fixed broadband speed of the global average is lower when compared to other case studies. As the data transfer speed decreases in the 4G and 3G case studies, we can see an increase in packet drop ratio for non-offloaded and MNO flows. If these flows are offloaded to the WiFi network, the packet delivery ratio rises to a better quality of service requirement.

\begin{figure}[ht]
\centering
\includegraphics[width=3.5in]{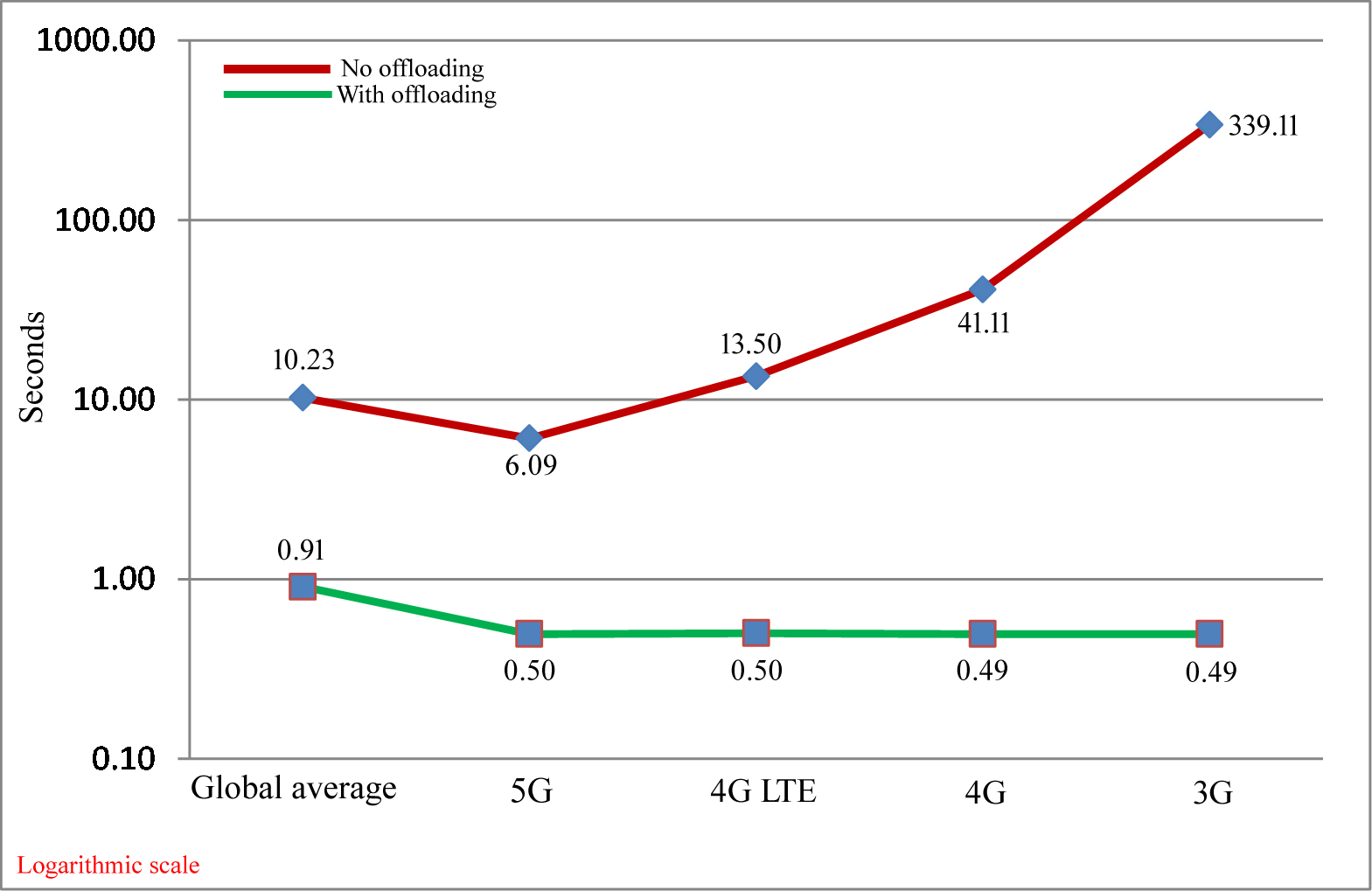}
\caption{Total Flow Duration Analysis For Offloading and No-offloading}
\label{fig:figure4}
\end{figure}

Fig. \ref{fig:figure4} exhibits the time taken to deliver a 500MB file, and 10 such total requests are made for each WiFi and MNO network. One request of transferring a 500MB file is then offloaded to a high-speed network. From this figure, it is evident that the time taken to transfer files is significantly reduced in the case of the offloaded flow. In the global average case study, the non-offloaded flow takes 10.23 seconds to transfer one file, whereas it is reduced to 0.91 seconds if the request is offloaded. Similarly, the graph shows a drastic reduction of delivery time in all other case studies. The longest time is taken by the 3G network which is 339.11 seconds to deliver the file, whereas it is reduced to only 0.49 seconds if this flow is offloaded to the WiFi network.

\begin{figure}[ht]
\centering
\includegraphics[width=3.5in]{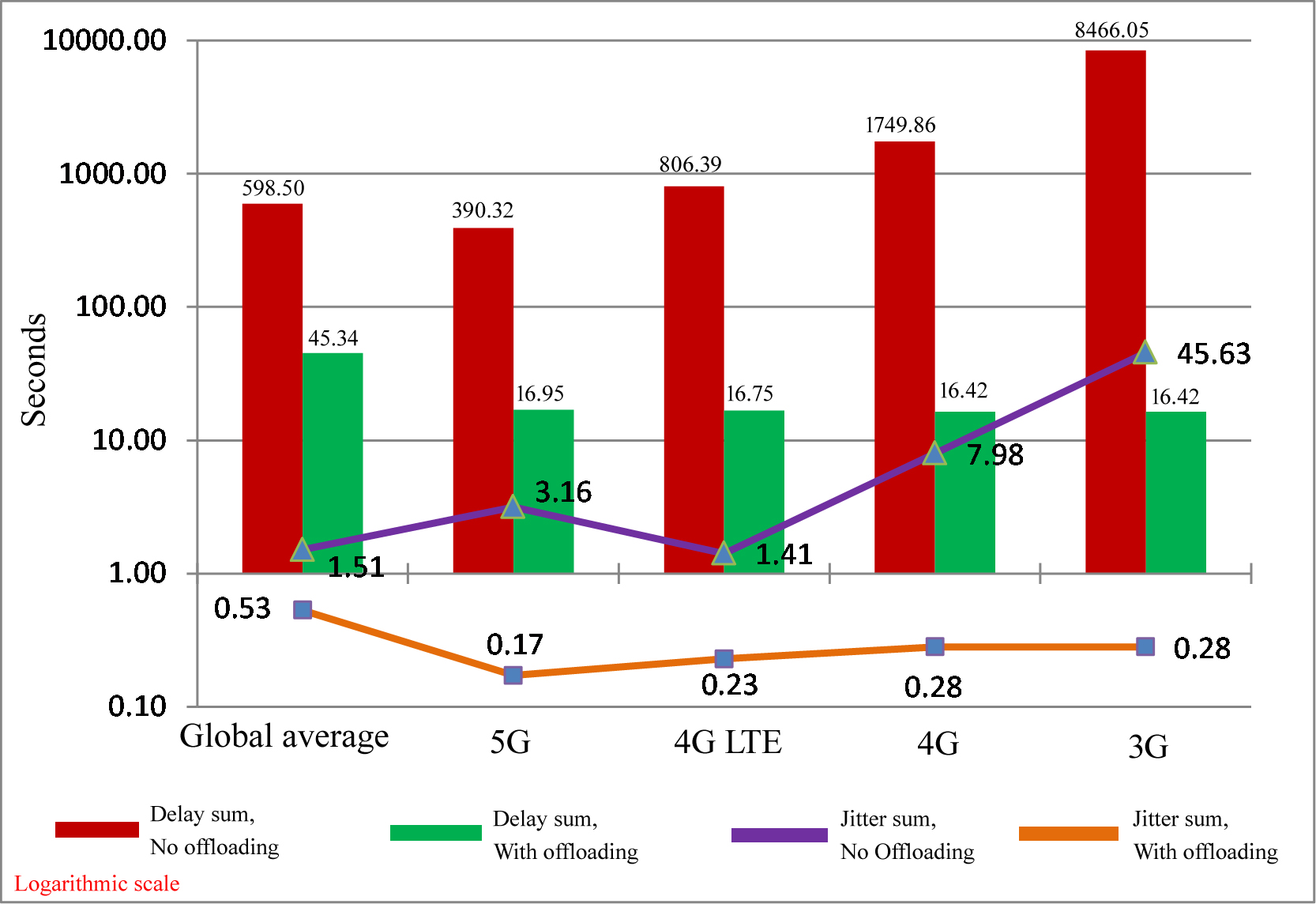}
\caption{Delay and Jitter Sum Analysis for Offloading and No-offloading}
\label{fig:figure5}
\end{figure}

Fig. \ref{fig:figure5} presents the analysis of delay sum and jitter sum for all case studies. The delay sum is the addition of all delays for each packet for the full duration of flow, whereas the jitter sum is the addition of all jitter for every packet for any particular flow. For the global average, the delay sum is calculated as 598.50 seconds. For other case studies such as 5G, 4G-LTE, 4G, and 3G it is computed as 390.32, 806.39, 1749.86, and 8466.05 respectively for the non-offloaded flows. If compared to the delay sum obtained by the offloaded flows, we can see a drastic reduction. The delay sums obtained for the offloaded flows are 45.34, 16.95, 16.75, 16.42, and 16.42 for global average, 5G, 4G-LTE, 4G and 3G.

Jitter sum is also improved in case of offloading, whereas it is high if no offloading is performed. The jitter sum of 1.51 seconds is reduced to 0.53 seconds for the global average case study if offloading is performed. It is reduced to 0.17 seconds from 3.16 for 5G, 0.23 seconds from 1.41 seconds for 4G-LTE, 0.28 seconds from 7.98 seconds for 4G, and impressive 0.28 seconds from 45.63 seconds for the 3G case study if offloading is performed. The enhanced packet delivery ratio, along with improved delivery time, reduced delay sum and reduced jitter sum, enhance the overall quality of experience provided by the MNOs to their subscribers. 

\section{Conclusions}
In this paper, an offloading framework is presented with its several advantages and infrastructural benefits. It can facilitate MNOs to allow their subscribers to benefit from third party operator high-speed infrastructure for a particular time period. To the best of our knowledge, there are no such mechanisms that exist as of now which can support such inter-organizational arrangements securely and efficiently. Our proposed framework deploys smart contracts for authentication, thereby allowing the subscriber to offload, and utilizes blockchain transactions to record and generate invoices. The MNO and private WiFi Provider both authenticate the subscriber and its mobile number to rule out any spamming of the system. All the transactions are verified by each participating entity, and new blocks are added using a PoA consensus mechanism to minimize the mining effort.   
 
The proposed framework was simulated using ns-3, and the Ethereum blockchain was integrated into the simulation environment using Docker containers. A total of five case studies i.e. global average speed, 5G, 4G-LTE, 4G, 3G to WiFi offloading were tested. The final analysis shows that offloading results in improved packet delivery ratios, reduced total flow duration, total delay, and total jitter. These parameters suggest that offloading can help in enhancing the end users quality of service experience. A present, Internet speeds vary geographically as well as amongst operators. A user has no choice but to switch operator if they need higher speeds, or services that cannot be provided by their operator. Our proposed offloading framework can be a great leap forward for subscribers who can enjoy higher bandwidth speeds on a on-demand basis.

In the future, challenges such as the scalability of blockchain transactions, simultaneous subscriber load, etc. can be analyzed. Time lag analysis of the high load of subscriber offloading requests can also be carried out. In addition, the same offloading framework can be investigated on automated switching of network traffic from high-congested channels to low-congested channels. This automated and agent-based framework could support load balancing and optimization of next-generation network infrastructure.

\end{document}